\documentclass[prd,preprint,superscriptaddress,showpacs,nofootinbib,%
tightenlines ]{revtex4}
\usepackage{epsfig}
\usepackage{amssymb}
\newcommand{\gA}{\stackrel{\circ}{g}_{\!A}}

\newcommand{\ben}{\begin{displaymath}}
\newcommand{\een}{\end{displaymath}}
\newcommand{\be}{\begin{equation}}
\newcommand{\ee}{\end{equation}}
\newcommand{\bea}{\begin{eqnarray}}
\newcommand{\eea}{\end{eqnarray}}
\begin{document}
%\preprint{MKPH-T-06-16}
\title{NN scattering in higher derivative formulation of baryon \\ chiral perturbation theory}
\author{D.~Djukanovic}
\affiliation{Institut f\"ur Kernphysik, Johannes
Gutenberg-Universit\"at, D-55099 Mainz, Germany}
\author{J.~Gegelia}
\affiliation{Institut f\"ur Kernphysik, Johannes
Gutenberg-Universit\"at, D-55099 Mainz, Germany}
\affiliation{Institute of High Energy Physics and Informatization,
Tbilisi State University, 0186 Tbilisi, Georgia}
\author{S.~Scherer}
\affiliation{Institut f\"ur Kernphysik, Johannes
Gutenberg-Universit\"at, D-55099 Mainz, Germany}
\author{M.~R.~Schindler}
\affiliation{Institut f\"ur Kernphysik, Johannes
Gutenberg-Universit\"at, D-55099 Mainz, Germany}
\date{Today}
\begin{abstract}
We consider a new approach to the nucleon-nucleon scattering problem
in the framework of the higher-derivative formulation of baryon
chiral perturbation theory. Starting with a Lorentz-invariant form
of the effective Lagrangian we work out a new symmetry-preserving
framework where the leading-order amplitude is calculated by solving
renormalizable equations and corrections are taken into account
perturbatively. Analogously to the KSW approach, the (leading)
renormalization scale dependence to any finite order is absorbed in
the redefinition of a finite number of parameters of the effective
potential at given order. On the other hand, analogously to
Weinberg's power counting, the one-pion-exchange potential is of
leading order and is treated non-perturbatively.

\end{abstract}

\maketitle

\section{\label{introduction}Introduction}

    The few-nucleon sector of baryon chiral perturbation theory
was first considered in the seminal papers by Weinberg
\cite{Weinberg:rz,Weinberg:um} which triggered an intensive
research activity starting with the pioneering work of
Ref.~\cite{Ordonez:1992xp}.
   In this approach for processes involving $N>1$ nucleons the
power counting is applied to the effective potential, which is
defined as the sum of all $N$-nucleon-irreducible diagrams.
  The scattering amplitudes are obtained by solving the
Lippmann-Schwinger (LS) or Schr\"odinger equation. For recent
reviews and references see e.g.
Refs.~\cite{Bedaque:2002mn,Epelbaum:2005pn}.

    The implementation of Weinberg's program encountered problems.
The NN potential of the effective field theory (EFT) is
non-renormalizable in the traditional sense already at leading
order. The iteration of the potential in the LS equation generates
divergent terms with structures which are not included in the
original potential. Therefore, the renormalization of the solution
to the LS equation requires the contributions of an infinite
number of higher-order counter-terms. Here, the infinite number
refers to both the loop and the chiral expansions.
   The freedom of choosing the finite parts of these counter-terms is compensated
by the running of the corresponding renormalized couplings.
   It has been argued that the coefficients of the divergent parts of the
counter-terms contributing in low-order calculations would set the
scale of the corresponding renormalized couplings.
   As a consequence, even if these couplings were natural at some value
of the renormalization scale, they would become unnaturally large
for slightly different values of this parameter.
   This problem (also analyzed non-perturbatively), in different variations,
has been addressed as the inconsistency of Weinberg's approach, and
an alternative power counting scheme, known as the KSW approach, has
been suggested
\cite{Kaplan:1996xu,Kaplan:1998tg,Kaplan:1998we,Savage:1998vh}. In
the KSW scheme the troublesome one-pion-exchange (OPE) potential is
shifted from leading order (LO) to next-to-leading order (NLO). The
leading-order equation becomes renormalizable, i.e. all divergences
can be absorbed in the redefinition of the parameters of the LO
potential, perturbatively as well as non-perturbatively. Moreover,
the LO equation is exactly solvable and dimensional regularization
can be applied. Corrections are treated perturbatively in the KSW
approach. This guarantees that {\it all} divergences are absorbed in
parameters contributing to the given order. As a consequence, the
KSW approach is free of ''consistency problems.'' Unfortunately this
scheme suffers from convergence problems. As the OPE potential is
shifted to NLO, it is treated perturbatively. It has been argued
that the perturbative treatment of pions in NN scattering poses
problems of convergence
\cite{Gegelia:1998ee,Cohen:1999ia,Gegelia:1999ja}. The results of
Ref.~\cite{Fleming:1999ee} made it explicit that perturbative series
do not converge in the KSW approach, at least in the triplet
channel.

As a next step, in Ref.~\cite{Beane:2001bc} a non-perturbative
investigation of the renormalization problem has been carried out
and an expansion of nuclear forces about the chiral limit has been
suggested. It has been argued that this expansion is formally
consistent and is equivalent to the KSW power counting in the
$^1S_0$ channel and to Weinberg's power counting in the $^3S_1 -
^3D_1$ coupled channels. Higher partial waves have been
investigated in Ref.~\cite{Nogga:2005hy} with the result that the
perturbative treatment of the OPE potential is not sufficient for
a finite number of partial waves, e.g. $^3P_0$ and $^3P_2 -
^3F_2$. On the other hand, the iterated OPE potential produces a
cutoff dependence in all waves where the tensor force is
attractive. This requires additional counter-terms in each of
these partial waves, clearly upsetting the power counting. The
authors of Ref.~\cite{Nogga:2005hy} conjecture that the ``mixture
of perturbative treatment of higher partial waves, resummation of
lower partial waves, and promotion of a finite number of
counter-terms is the most consistent approach'' to the chiral
perturbation theory for nuclear systems while sub-leading
interactions are treated perturbatively.

\medskip
\medskip

There exists an alternative point of view that the ''consistency
problem'' of Weinberg's approach is irrelevant and one can perform
the calculations in cutoff effective field theory (EFT) by suitably
choosing the cutoff parameter. According to this point of view,
taking large values of the cutoff parameter without including the
contributions of all (perturbatively) relevant counter-terms is not
a legitimate procedure. Instead one has to choose the cutoff
parameter of the order of the relevant large scale, presumably the
mass of the $\rho$ meson
\cite{Lepage:1997cs,Gegelia:gn,Gegelia:1998iu,Park:1998cu,Lepage:1999kt,Epelbaum:2004fk,Gegelia:2004pz,Epelbaum:2006pt}.

\medskip

In the present work we develop a new approach which combines aspects
of both points of view outlined above. In particular, in our new
framework the leading-order amplitude is calculated by solving
integral equations and corrections are taken into account
perturbatively. Analogously to the KSW approach, the (leading)
renormalization scale dependence to any given finite order is
absorbed in the redefinition of a finite number of parameters of the
effective potential at the given order. On the other hand,
analogously to Weinberg's scheme, the OPE potential is of leading
order and is treated non-perturbatively.

%The applied regularization scheme preserves all underlying
%symmetries to the given order of accuracy and the self-consistency
%of the power counting is explicit in this new formulation.
In the two-nucleon sector the path integral for the generating
functional of the Green's functions is not Gaussian and therefore
the standard heavy-baryon reduction can not be carried out. While
the non-relativistic effective Lagrangian can be fixed by matching
to the (explicitly) Lorentz-invariant Lagrangian, we prefer to work
in terms of the latter. We start with the higher-derivative
formulation of the manifestly Lorentz-invariant Lagrangian of
Ref.~\cite{Djukanovic:2004px} and perform the non-relativistic
reduction at the level of Feynman diagrams.

Our work is organized as follows: In Sec.~\ref{effective_Lagrangian}
the effective Lagrangian in its manifestly Lorentz-invariant form
and field redefinitions leading to the higher-derivative formulation
are considered. Section \ref{Renormalization} addresses the issues
of renormalization and power counting. In Sec.~\ref{equations} the
NN scattering equations are derived and analyzed. A summary is given
in Sec.~\ref{conclusions}.

%In the appendix the perturbative renormalizability of our LO
%equation is demonstrated for one- and two-loop diagrams.

\section{The effective Lagrangian}
\label{effective_Lagrangian}

    In this section we consider the effective Lagrangian needed
for the purposes of this work.
  The standard effective Lagrangian is organized in a
derivative and quark-mass expansion and consists of the sum of the
purely mesonic, the $\pi {\rm N}$, ${\rm NN}$, $\ldots$ parts,
\begin{equation}
{\cal L}_{\rm eff}={\cal L}_{\pi}+{\cal L}_{\pi {\rm N}}+{\cal
L}_{\rm NN}+\cdots. \label{inlagr}
\end{equation}
The lowest-order mesonic Lagrangian reads
\cite{Gasser:1984yg}
\begin{equation}\label{piLag}
\label{l2} {\cal L}_2=\frac{F^2}{4}\mbox{Tr}\left[D_\mu U \left(
D^\mu U\right)^\dagger\right] +\frac{F^2}{4}\mbox{Tr} \left( \chi
U^\dagger+ U\chi^\dagger \right),
\end{equation}
where $U$ is a unimodular unitary $(2\times 2)$ matrix containing
the pion fields. The covariant derivative is defined as
$$
D_\mu U=\partial_\mu U-i r_\mu U+i U l_\mu,
$$
and
$$r_\mu=v_\mu+a_\mu, \ \ \ l_\mu=v_\mu-a_\mu, \ \ \ \chi =2 B (s+i p). $$
Here, $v_\mu$, $a_\mu$, $s$, and $p$ are external vector,
axial-vector, scalar, and pseudo-scalar sources, respectively.
   In Eq.\ (\ref{l2}), $F$ denotes the pion-decay constant in the chiral
limit: $F_\pi=F[1+{\cal O}(\hat{m})]=92.4$ MeV.
   We consider the isospin-symmetric limit $m_u=m_d=\hat{m}$,
and the lowest-order expression for the squared pion mass is
$M^2=2 B \hat{m}$, where $B$ is related to the quark condensate
$\langle \bar{q} q\rangle_0$ in the chiral limit
\cite{Gasser:1984yg}.

  The nucleon field is described with two four-component Dirac
fields,
\begin{displaymath}
\Psi=\left(\begin{array}{c}p\\n\end{array}\right)\,,
\end{displaymath}
and the lowest-order Lagrangian of the one-nucleon sector reads
\cite{Gasser:1988rb}
\begin{equation}
{\cal L}_{\pi {\rm N}}^{(1)}=\bar \Psi \left( i\gamma_\mu D^\mu -m
+\frac{1}{2} \gA\gamma_\mu \gamma_5 u^\mu\right) \Psi,
\label{lolagr}
\end{equation}
   where $D_\mu\Psi = (\partial_\mu +\Gamma_\mu-i v^{(s)}_\mu)\Psi $
denotes the covariant derivative with the iso-scalar vector source
$v_\mu^{(s)}$, and the quantities $u$, $u_\mu$, $\Gamma_\mu$,  and
$\chi_{\pm}$ (and their derivatives) are defined as
$$
u^2=U,\quad u_\mu =iu^{\dagger} D_\mu U u^{\dagger},\quad
\Gamma_{\mu}=\frac{1}{2}\left[u^\dagger\partial_{\mu}u
+u\partial_{\mu}u^\dagger- i(u^{\dagger}r_{\mu}u+u
l_{\mu}u^{\dagger})\right],
$$
$$
\chi_{\pm}=u^{\dagger}\chi u^\dagger\pm u\chi^\dagger u.
$$
     In Eq.~(\ref{lolagr}), $m$ and $\gA$ stand for the chiral limit of
the physical nucleon mass and the axial-vector coupling constant,
respectively.

    The most general chirally-invariant leading-order NN Lagrangian reads (for the power counting see below)
\begin{eqnarray}
{\cal L}_{\rm NN}&=& \sum_{a=0}^3\,\Biggl[ C_S^a \ \bar\Psi
\tau^a\Psi \ \bar\Psi\tau^a \Psi +C_A^a \ \bar\Psi \tau^a \gamma_5
\Psi \ \bar\Psi\tau^a \gamma_5  \Psi + C_V^a \ \bar\Psi\tau^a
\gamma_\mu \Psi \ \bar\Psi\tau^a \gamma^\mu \Psi \nonumber \\
   &&+\, C_{AV}^a \ \bar\Psi\tau^a\gamma_5 \gamma_\mu\Psi \
\bar\Psi\tau^a\gamma_5\gamma^\mu\Psi + C_{T}^a \
\bar\Psi\tau^a\sigma_{\mu\nu} \Psi \
\bar\Psi\tau^a\sigma^{\mu\nu}\Psi\Biggr] \,, \label{NNLagrdreg}
\end{eqnarray}
where $\tau^0$ is the unit (isospin) matrix and $\tau^i$ ($i=1,2,3$)
are (isospin) Pauli matrices and $C_I^1=C_I^2=C_I^3$ for all $I$.
Note that the structure $\bar\Psi \gamma_5\Psi$ is assigned  leading
order.

   The most general effective Lagrangian
includes {\it all} terms which are consistent with the symmetries
of the underlying fundamental theory, i.e. QCD. The above given
"canonical" form of the effective Lagrangian is obtained from the
most general form by using suitable field redefinitions
\cite{Scherer:1994wi}. This is done for convenience. Any two forms
of the effective Lagrangian obtained from each other by using
field redefinitions are completely equivalent when physical
quantities are considered. Note that the coupling constants
corresponding to the same operator structures in two forms of the
effective Lagrangian are in general not equal
\cite{Belkov:1994qg}. Switching from one formulation to the other,
one needs to take into account the new contributions (in general
in all operator structures) generated by field transformations.
There is an infinite number of possible field transformations and
correspondingly an infinite number of forms of the effective
Lagrangian. While all these forms lead to the same expressions for
physical quantities (with certain care), in practical
(perturbative) calculations one form of the effective Lagrangian
could be considerably more convenient than the other.

    Starting with the above given "canonical" effective Lagrangian
one encounters ultraviolet divergences in the equations for the
few-nucleon scattering amplitudes. This problem can be handled by
considering a nucleon propagator with improved ultraviolet
behavior \cite{Djukanovic:2004px}. To that end we employ a
"non-canonical" form of the effective Lagrangian generated by
using suitable field transformations.
    To obtain the modified nucleon propagator we
perform the following field transformations
\begin{eqnarray}
\label{fredefinition}
  \Psi &\rightarrow&  \Phi \left( L^2, \, -\overrightarrow{D}^2 \right)
\Psi\,,\nonumber\\
  \bar{\Psi} &\rightarrow& \bar{\Psi} \,\Phi \left( L^2, \, - \overleftarrow{D}^2\right)\,,
\end{eqnarray}
where $\bar\Psi\, \overleftarrow D^2 =\left(
\overrightarrow D^2 \Psi\right)^\dagger \gamma^0 $ and obtain
\begin{eqnarray}
{\cal L}_{\pi {\rm N}}^{(1)'}=\bar \Psi \,\Phi \left( L^2, \,
-\overleftarrow{D}^2\right)\,\left( i\gamma_\mu D^\mu -m
+\frac{1}{2} \gA\gamma_\mu \gamma_5 u^\mu\right) \Phi\left( L^2, \,
-\overrightarrow{D}^2\right) \Psi\,.
 \label{LOLprime}
\end{eqnarray}
In Eq.~(\ref{LOLprime}) $L$ is a free parameter,
$\overrightarrow{D}$ stands for the spacial components of the
covariant derivative and $\Phi$ is some appropriately chosen real
function. Possible choices of $\Phi$ are, e.g.,
\begin{equation}
\Phi\left( L^2,X\right)=\left( 1 + \frac{1}{L^2} \, X
\right)^{N_\Psi} \ {\rm or} \ \Phi\left( L^2,X\right)=\exp \left(
\frac{1}{L^2} \, X \right)^{N_\Psi}\,. \label{Ffactors}
\end{equation}
    The Feynman propagator corresponding to the Lagrangian of
Eq.~(\ref{LOLprime}) reads
\begin{equation}
i\,S_F(p)=\frac{i}{p \hspace{-.45em}/\hspace{.1em}
-m+i\,\epsilon}\,F(L^2,\vec p\,{}^2)\,, \label{fp}
\end{equation}
with
\begin{equation}
F(L^2,\vec p\,{}^2)=\frac{L^{4\,N_\Psi}}{\left( L^2+\vec
p^2\right)^{2\,N_\Psi}}\, \ {\rm and} \ F(L^2,\vec p\,{}^2)=\exp
\left\{ -2 \left( \frac{\vec p\,^2 }{L^2}\right)^{N_\Psi}\right\}\,,
\label{ff}
\end{equation}
respectively.

    Depending on the order of the performed calculations
we choose the modified propagator such that all loop integrations
over the momenta of the intermediate purely nucleonic states are
rendered finite. We apply dimensional regularization to the
resulting effective theory, expand the regularized diagrams in
powers of $n-4$ and subtract $1/(n-4)$ pole-terms, where $n$ is the
space-time dimension. There is a finite number of pole-terms to any
given (finite) order in the chiral expansion of physical quantities
in the vacuum and one-nucleon sectors, and the effective potentials
in the two-nucleon sector. No further divergences occur (for finite
parameters $L$) neither in the vacuum and one-nucleon sector nor in
the equations of the two-nucleon sector. Therefore, we can take
$n=4$ in the equations of the two-nucleon sector after renormalizing
the effective potential and the corrections to the two-nucleon
propagator. Note that the field redefinition of
Eq.~(\ref{fredefinition}) introduces additional interaction terms in
the effective Lagrangian. Their contributions are important for
preserving the symmetries of the theory. In practice it is
convenient to choose the exponential form of the field redefinition
with sufficiently large $N_\Psi$ \cite{Epelbaum:2005pn} where the
additional interaction terms start contributing at orders higher
than the order of accuracy of the performed calculations.

The complete dependence of the physical quantities on the parameter
$L$ can be absorbed in the redefinition of (an infinite number of)
coupling constants. Therefore different choices of the numerical
value of this parameter correspond to different renormalization
schemes. Note that the applied regularization scheme is the
dimensional regularization. The parameter $L$ should not be confused
with the cutoff parameter and it does not have to be taken to
infinity.

\section{Power counting}
\label{Renormalization}

It is understood that the power counting specified below is
satisfied by renormalized diagrams, provided the renormalization
points are chosen appropriately. In particular, following
Ref.~\cite{Weinberg:um} one should choose the renormalization
points of the order of small external momenta.

\subsection{Effective potential}

Interaction terms, generated by the one-nucleon sector of the
effective Lagrangian \cite{Fettes:2000gb}  are assigned the standard
orders of BChPT \cite{Ecker:1995gg}. In the two-nucleon sector each
derivative acting on the pion field as well as each factor of the
pion mass is assigned order $q^1$, where $q$ denotes a small
quantity like the pion mass or a nucleon three momentum. Contracted
derivatives acting on nucleon fields lead to scalar products of
large nucleon momenta, e.g. $p_1 \cdot p_2$. We can rewrite
\begin{equation}\label{PCsaclarpro}
   2 p_1 \cdot p_2 = (p_1^2-m^2)+(p_2^2-m^2)-(p_1-p_2)^2+2m^2.
\end{equation}
As in the one-nucleon sector, $(p_i^2-m^2)$ is counted as $q^1$,
while the difference $(p_1-p_2)^2$ is of order $q^2$. We arrange the
Lagrangian such that terms involving derivatives are accompanied by
interaction terms where each pair of contracted derivatives is
replaced with a factor of ''$+m^2$'' or ''$- m^2$''. Therefore each
pair of contracted derivatives acting on nucleon fields is assigned
order $q^1$ (we choose the renormalization scheme such that this
power counting also holds for loop diagrams.) For example, the term
$\bar\Psi D_\mu \Psi \bar\Psi D^\mu \Psi$ is rewritten as
\begin{equation}
\left(\bar\Psi D_\mu \Psi \bar\Psi D^\mu \Psi +
m^2\,\bar\Psi\Psi\bar\Psi\Psi\right)-m^2\,\bar\Psi\Psi\bar\Psi\Psi\,,
\label{NNLexample}
\end{equation}
where the first and second terms are part of the NLO and LO NN
Lagrangian, respectively.

We assign to each diagram contributing to the effective potential a
chiral order $D$ determined as follows. Vertices from the $k$-th
order Lagrangian count as order $q^k$, a loop integration in $n$
dimensions as $q^n$, each pion propagator as $q^{-2}$ and each
nucleon propagator as $q^{-1}$, respectively. Note that the terms of
the $k$-th order NN Lagrangian can only generate contributions to
effective potential of order $k$ and higher. To any finite order
there is only a finite number of terms in the two-nucleon sector of
the effective Lagrangian. Correspondingly the effective potential
contains a finite number of terms to any finite order. We drop
nucleon loops, including their contributions in the redefinition of
the coupling constants of the effective Lagrangian.

\subsection{Two-nucleon propagator}
We rewrite the fermion propagator as
\begin{eqnarray}
S_F(p) &=& \frac{p\hspace{-.45em}/\hspace{.1em}+m}{p^2-m^2+i\,\epsilon}\, F(L^2,\vec p\,{}^2)\nonumber\\
&=&\frac{m\,P_+ F(L^2,\vec p\,{}^2) }{\sqrt{\vec p^2+m^2}\,\left(
p_0-\sqrt{\vec p^2+m^2}+i\,\epsilon\right)}-\frac{m\,P_+ F(L^2,\vec
p\,{}^2)}{\sqrt{\vec p^2+m^2}\,\left( p_0+\sqrt{\vec
p^2+m^2}-i\,\epsilon\right)}\nonumber\\
  &&+ \frac{\left(p
\hspace{-.45em}/\hspace{.1em}-m\,v \hspace{-.45em}/\hspace{.1em}
\right)F(L^2,\vec p\,{}^2)}{2\,\sqrt{\vec p^2+m^2}\,\left(
p_0-\sqrt{\vec p^2+m^2}+i\,\epsilon\right)}-\frac{\left(p
\hspace{-.45em}/\hspace{.1em}-m\,v
\hspace{-.45em}/\hspace{.1em}\right)F(L^2,\vec p\,{}^2)
}{2\,\sqrt{\vec p^2+m^2}\,\left(p_0+\sqrt{\vec
p^2+m^2}-i\,\epsilon\right)}\nonumber\\
&=& S_F^{(1)}(p)+S_F^{(2)}(p)+S_F^{(3)}(p)+S_F^{(4)}(p)
\,,\label{Sfpexpanded}
\end{eqnarray}
where the projector $P_+$ is defined as
\begin{equation}\label{PPlus}
P_+ = \frac{1+v \hspace{-.55em}/\hspace{.1em}}{2}\,,
\end{equation}
with $v=(1,0,0,0)$. The poles in $S_F^{(1)}$ and $S_F^{(3)}$
correspond to particle states, and those in $S_F^{(2)}$ and
$S_F^{(4)}$ to anti-particles, respectively. We assign order
$q^{-1}$ to $S_F^{(1)}$,  order $q^0$ to the second and third
terms $S_F^{(2)}$ and $S_F^{(3)}$, and order $q^1$ to $S_F^{(4)}$.

The undressed two-nucleon propagator is defined as the product of
two undressed one-nucleon propagators. Inserting the expansion of
Eq.~(\ref{Sfpexpanded}) we obtain a sum of terms of different
orders. For nucleon-nucleon scattering the products of $S_F^{(1)}$
and $S_F^{(3)}$ are enhanced in purely two-nucleon intermediate
states \cite{Weinberg:rz,Weinberg:um}. Therefore, we assign the
following orders to the various terms in the undressed two-nucleon
propagator:
\renewcommand{\arraystretch}{1.5}
\begin{equation}
\begin{array}{llll}
S_F^{(1)} S_F^{(1)} \sim q^{-4}\,, &&& S_F^{(1)} S_F^{(3)} \sim
q^{-3}\,,\\ S_F^{(2)} S_F^{(2)} \sim q^{0}\,, &&& S_F^{(2)}
S_F^{(4)} \sim q^{1}\,,\\
S_F^{(3)} S_F^{(3)} \sim q^{-2}\,, &&& S_F^{(4)} S_F^{(4)} \sim
q^{2}\,,\\
S_F^{(1)} S_F^{(2)} \sim q^{-1}\,, &&& S_F^{(1)} S_F^{(4)} \sim
q^{0}\,,\\
S_F^{(2)} S_F^{(3)} \sim q^{0}\,, &&& S_F^{(3)} S_F^{(4)}\sim
q^{1}\,.\\
\label{twoNprop}
\end{array}
\end{equation}
The first term in Eq.~(\ref{twoNprop}) defines the leading-order
two-nucleon propagator $G_0$,
\begin{equation}\label{G0Def}
    G_0=-i\,S_F^{(1)} S_F^{(1)}\,.
\end{equation}
The last four terms in Eq.~(\ref{twoNprop}) possess poles on the
same side of the real axis in the complex plane of the zeroth
component of the integration momenta. Therefore, if the effective
potential does not contain singularities in the complex energy
plane, they give vanishing contributions.

\section{NN scattering equation}
\label{equations}

Defining the two-nucleon-irreducible diagrams as the effective
potential $V$ the equation for the amplitude (amputated Green's
function) reads
\begin{eqnarray}
T_{\lambda\sigma,\mu\nu}\left(
p_3,p_4,p_1,p_2\right)&=&V_{\lambda\sigma,\mu\nu}\left(
p_3,p_4,p_1,p_2\right)\nonumber\\
   &&- i\, \int \frac{d^4 k}{(2\,\pi)^4} \ V_{\lambda\sigma,
\alpha\gamma}\left( p_3,p_4,p_3-k,p_4+k\right) \nonumber\\
   &&\times S_{F,\alpha\beta} (p_3-k)\, S_{F,\gamma\delta} (p_4+k) \
T_{\beta\delta,\mu\nu}\left( p_3-k,p_4+k,p_1,p_2\right)\,,
\label{MeqFull}
\end{eqnarray}
where the Greek indices labeling the components of the Dirac fermion
field run from 1 to 4. The corresponding physical amplitude is
obtained as
\begin{equation}
\sqrt{Z_\psi(p_3) }\,\sqrt{Z_\psi(p_4) }\, \bar u_\lambda
(p_3)\bar u_\sigma (p_4)\,T_{\lambda\sigma,\mu\nu}\left(
p_3,p_4,p_1,p_2\right)\,u_\mu (p_1)u_\nu (p_2)\,\sqrt{Z_\psi(p_1)
}\,\sqrt{Z_\psi(p_2) }\,, \label{LSZtosh}
\end{equation}
where $Z_\psi$ is the residue of the nucleon propagator. The
residue corresponding to the propagator of Eqs.~(\ref{fp}) and
(\ref{Sfpexpanded}) is
\begin{equation}
Z_\psi(p) =F(L^2,\vec p\,{}^2)\,. \label{residue}
\end{equation}
For convenience in the following derivations we write
Eq.~(\ref{MeqFull}) symbolically as
\begin{equation}
T_{\lambda\sigma,\mu\nu} = V_{\lambda\sigma,\mu\nu} +
V_{\lambda\sigma, \alpha\gamma}\,G_{\alpha\beta,\gamma\delta}\
T_{\beta\delta,\mu\nu} \,, \nonumber
\end{equation}
or
\begin{equation}
T = V + V \,G \ T \,.\label{EqForGF}
\end{equation}
Next, we expand $T$, $V$, and $G$ in small parameters,
\begin{eqnarray}
G & = & G_0 +G_1 +G_2 +\cdots\,, \label{Gexpanded} \\
V & = & V_0 +V_1 +V_2 +\cdots\,, \label{Vxpanded}
\\
T & = & T_0 +T_1 +T_2 +\cdots\,, \label{Texpanded}
\end{eqnarray}
where the quantities $X_i$ ($X=T,G,V$) are of order ${\cal O}(q^i)$.
We then substitute the expansions in Eq.~(\ref{EqForGF}), and solve
$T$ order by order. At leading order we obtain
\begin{equation}
T_0 = V_0 + V_0\, G_0\, T_0\,. \label{LOgeq}
\end{equation}
To determine the physical amplitude order by order
%we rewrite the
%solutions to the Dirac equation as
%\begin{eqnarray}
%u(p) & = & \frac{1+v \hspace{-.45em}/\hspace{.1em}}{2}
%u(p)+\frac{1-v \hspace{-.45em}/\hspace{.1em}}{2} u(p)=\left( 1+
%\frac{p \hspace{-.45em}/\hspace{.1em}-p\cdot v}{m+p\cdot v}\right)
%\frac{1+v \hspace{-.45em}/\hspace{.1em}}{2} u(p),
%\\
%\bar u(p) & = & \bar u(p) \frac{1+v \hspace{-.45em}/\hspace{.1em}
%}{2}+\bar u(p)\frac{1-v \hspace{-.45em}/\hspace{.1em}}{2}=\bar
%u(p) \frac{1+v \hspace{-.45em}/\hspace{.1em}}{2} \left( 1+ \frac{p
%\hspace{-.45em}/\hspace{.1em}-p\cdot v}{m+p\cdot v}\right),
%\end{eqnarray}
%and
we expand the solutions to the Dirac equation in small quantities
\begin{eqnarray}
u & = & u_0 + u_1 + u_2 +\cdots = P_+\, u(p)+ \frac{p
\hspace{-.45em}/\hspace{.1em} -p\cdot v}{2\,m} \,P_+ \,u(p)+\cdots
\,,\nonumber
\\
\bar u & = &  \bar u_0 +\bar u_1 +\bar u_2 +\cdots = \bar u(p)\,P_+
+\bar u(p) \,P_+ \frac{p \hspace{-.45em}/\hspace{.1em} -p\cdot
v}{2\,m} + \cdots \,. \label{deqsolexpanded}
\end{eqnarray}
%where
%\begin{eqnarray}
%u_0(p) &=& P_+\, u(p)\,, \nonumber \\ \bar u_0(p) & = & \bar u(p)\,P_+\,,\nonumber\\
%u_1(p)
%   & = & \frac{p \hspace{-.45em}/\hspace{.1em}
%-p\cdot v}{2\,m}
%\,P_+ \,u(p)\,, \nonumber\\
%\bar u_1(p)
%   & = & \bar u(p) \,P_+ \frac{p
%\hspace{-.45em}/\hspace{.1em} -p\cdot
%v}{2\,m}\,,\nonumber\\
%   &\vdots& \nonumber\\
%%&& \cdots \ \ {\rm etc}\,.
%\label{spellout}
%\end{eqnarray}
Substituting Eqs.~(\ref{Texpanded}) and (\ref{deqsolexpanded}) in
Eq.~(\ref{LSZtosh}) we obtain the expansion of the physical
amplitude as
\begin{eqnarray}
&& \sqrt{Z_\psi}\,\sqrt{Z_\psi}\, \bar u\bar u\,T\,u u
\sqrt{Z_\psi}\,\sqrt{Z_\psi}\, = \sqrt{Z_\psi}\,\sqrt{Z_\psi}\,
\left[ \bar u_0\bar u_0\,T_0\,u_0 u_0+ \bar u_1\bar u_0\,T_0\,u_0
u_0+ \bar u_0\bar
u_1\,T_0\,u_0 u_0\right.\nonumber\\
&& \hspace{7em}\left.+\bar u_0\bar u_0\,T_1\,u_0 u_0+\bar u_0\bar
u_0\,T_0\,u_1 u_0+\bar u_0\bar u_0\,T_0\,u_0
u_1+\cdots\right]\,\sqrt{Z_\psi}\,\sqrt{Z_\psi}\,.
\label{LSZtoshExpand}
\end{eqnarray}
Taking into account Eq.~(\ref{deqsolexpanded}) we see that to
calculate the leading-order on-mass-shell amplitude we need the
quantity $\tilde T_0 = P_+ P_+\, T_0 \,P_+ P_+$. Applying the
projectors $P_+$ to Eq.~(\ref{LOgeq}) and using (see
Eq.~(\ref{G0Def}))
%\begin{eqnarray}
%G_{0; \alpha\beta,\mu\nu} &=& G_{0; \alpha\beta_1,\mu\nu_1} P_{+ ;
%\beta_1\beta} P_{+;\nu_1\nu} =P_{+;\alpha\alpha_1}
%P_{+;\beta \beta_1} G_{0; \alpha_1\beta_1,\mu\nu}\nonumber\\
%&=& P_{+ ; \alpha\alpha_1} P_{+ ; \beta\beta_1} G_{0;
%\alpha_1\beta_1,\mu_1\nu_1} P_{+;\mu_1\mu} P_{+;\nu_1\nu}\,
%\label{Gsubst}
%\end{eqnarray}
%or
\begin{equation}\label{GsubstShort}
    G_0=G_0 P_+ P_+=P_+ P_+ G_0=P_+ P_+ G_0 P_+ P_+\,
\end{equation}
we obtain
\begin{equation}
P_+ P_+\, T_0 \,P_+ P_+= P_+ P_+\, V_0 \,P_+ P_+ +P_+ P_+\, V_0
\,P_+ P_+\, G_0\,P_+ P_+\, T_0 \,P_+ P_+\,, \label{LOgeqP}
\end{equation}
which is an integral equation for the projected amplitude $\tilde
T_0$,
\begin{equation}
\tilde T_0= \tilde V_0+\tilde V_0\, G_0\,\tilde T_0 \,,
\label{LOgeqPTilde}
\end{equation}
where $\tilde V_0=P_+ P_+\, V_0 \,P_+ P_+$.

\medskip
To calculate $T_0$ from the solution to Eq.~(\ref{LOgeqPTilde}) we
rewrite Eq.~(\ref{LOgeq}) as
\begin{equation}
T_0 = V_0 + V_0\, G_0\, V_0 + V_0\, G_0\, T_0\, G_0\, V_0\,,
\label{LOgeq2}
\end{equation}
which, using Eq.~(\ref{GsubstShort}), takes the form
\begin{equation}
T_0 = V_0 + V_0\, G_0\, V_0 + V_0\, G_0\,\tilde T_0 \, G_0
\,V_0\,. \label{LOgeq2tilde}
\end{equation}

The leading-order amplitude $T_0$ calculated from
Eq.~(\ref{LOgeq2tilde}) can then be used in calculations of the
next-to-leading-order amplitude $T_1$. The equation for $T_1$
\begin{equation}
T_1=V_1+ V_1\,G_0\, T_0+ V_0\,G_1\,T_0+ V_0\,G_0\, T_1
\label{NLOeqS}
\end{equation}
can be solved exactly. The solution reads
\begin{equation}
T_1=V_1+ T_0\,G_0\, V_1+ V_1\,G_0\, T_0+ T_0\,G_0\,V_1\,G_0\,T_0+
T_0\,G_1\, T_0\,. \label{NLOeqsolutionS}
\end{equation}

Analogously to the above case the leading-order amplitude $T_0$
and the next-to-leading-order amplitude $T_1$ can be used to
calculate the next-to-next-to-leading-order amplitude $T_2$ etc.

\subsection{\label{LOcontactinteraction}Leading-order equation}

   In the following, we will consider the scattering amplitude in the
center-of-mass frame, where $p_1=(\sqrt{m^2+\vec{p}\,^2},\vec
p\,)$, $p_2 =(\sqrt{m^2+\vec{p}\,^2},-\vec p\,)$,
$p_3=(\sqrt{m^2+\vec{p\,'}\,^2},\vec p\,'\,)$, $p_4
=(\sqrt{m^2+\vec{p\,'}\,^2},-\vec p\,'\,)$.
   Let $P=p_1+p_2=(2p_0,\vec 0)$ denote the total four-momentum of the scattered
nucleons where $p_0=\sqrt{\vec{p}\,^2+m^2}$.

The leading-order effective potential consists of the contact
interaction part and the one-pion-exchange diagram. The
leading-order contact interaction potential reads
\begin{eqnarray}
i V_{0, C;\lambda\sigma ,\mu \nu}^{i_3 i_4,i_1 i_2} & = & 2\, i
\sum_{a=0}^3 \,\Biggl\{ C_S^a \left(
\delta_{\lambda\nu}\delta_{\sigma\mu}\tau^{a}_{i_3 i_2}\tau^{a}_{i_4
i_1} -\delta_{\lambda \mu}\delta_{ \sigma \nu}\tau^{a}_{i_3
i_1}\tau^{a}_{i_4 i_2}\right)
\nonumber\\
  &&+ C_A^a \left( \gamma_{5; \lambda \nu}\gamma_{5; \sigma \mu}\tau^{a}_{i_3
i_2}\tau^{a}_{i_4 i_1} -\gamma_{5; \lambda \mu}\gamma_{5; \sigma
\nu}\tau^{a}_{i_3
i_1}\tau^{a}_{i_4 i_2}\right)\nonumber\\
  &&+ C_V^a \left( \gamma^\alpha_{\lambda \nu}\gamma_{\alpha; \sigma \mu}\tau^{a}_{i_3
i_2}\tau^{a}_{i_4 i_1} -\gamma^\alpha_{\lambda
\mu}\gamma_{\alpha;\sigma \nu}\tau^{a}_{i_3
i_1}\tau^{a}_{i_4 i_2}\right) \nonumber\\
  &&+ C_{AV}^a \left[ \left( \gamma_5\gamma_\alpha\right)_{\lambda
\nu}\left(\gamma_5\gamma^\alpha\right)_{\sigma \mu}\tau^{a}_{i_3
i_2}\tau^{a}_{i_4 i_1} -\left( \gamma_5\gamma_\alpha\right)_{\lambda
\mu}\left( \gamma_5\gamma^\alpha\right)_{\sigma \nu}\tau^{a}_{i_3
i_1}\tau^{a}_{i_4 i_2}\right] \nonumber\\
  &&+ C_T^a \left[ \left( \sigma^{\alpha\beta}\right)_{\lambda \nu}\left(
\sigma_{\alpha\beta}\right)_{\sigma \mu}\tau^{a}_{i_3
i_2}\tau^{a}_{i_4 i_1} -\left( \sigma^{\alpha\beta}\right)_{\lambda
\mu}\left( \sigma_{\alpha\beta}\right)_{\sigma \nu}\tau^{a}_{i_3
i_1}\tau^{a}_{i_4 i_2}\right]\Biggr\} . \label{NNContactpot}
\end{eqnarray}
The projected potential is
\begin{eqnarray}
i \tilde V_{0, C;l s,m n}^{i_3 i_4,i_1 i_2} & = & 2 \, i
\sum_{a=0}^3 \, \Biggl\{\left( C_S^a + C_V^a\right) \left( \delta_{l
n}\delta_{s m}\tau^{a}_{i_3 i_2}\tau^{a}_{i_4 i_1} -\delta_{l
m}\delta_{ s n}\tau^{a}_{i_3
i_1}\tau^{a}_{i_4 i_2}\right)\nonumber\\
  &&- \left( C_{AV}^a+2 C_T^a\right)\,\sum_{b=1}^3 \left[
\sigma^b_{l n}\sigma^b_{s m}\tau^{a}_{i_3 i_2}\tau^{a}_{i_4 i_1} -
\sigma^b_{l m} \sigma^b_{s n}\tau^{a}_{i_3 i_1}\tau^{a}_{i_4
i_2}\right]\Biggr\}\,, \label{NNContactpotProjected}
\end{eqnarray}
with $l, s, m, n =1,2$. Note that the pseudoscalar piece of the
leading-order NN Lagrangian starts contributing at higher orders.

The covariant form of the one-pion-exchange potential reads
\begin{eqnarray}
i \, V^{i_3 i_4,i_1 i_2}_{0,OPE; \lambda \sigma ,\mu \nu}  \left(
p_3,p_4,p_1,p_2 \right) &=& -\sum_{a=1}^3 \, i\,
\frac{g_A^2}{4\,F^2}\,\frac{\left[ \gamma_5 \left( p
\hspace{-.45em}/\hspace{.1em}_1- p
\hspace{-.45em}/\hspace{.1em}_3\right)\right]_{\lambda\mu}\left[
\gamma_5 \left( p \hspace{-.45em}/\hspace{.1em}_1-p
\hspace{-.45em}/\hspace{.1em}_3\right)\right]_{\sigma\nu}}{\left(
p_1-p_3\right)^2-M^2}\
\tau^{a}_{i_3 i_1}\tau^{a}_{i_4 i_2}\nonumber\\
&+& \sum_{a=1}^3 \,i\, \frac{g_A^2}{4\,F^2}\,\frac{\left[ \gamma_5
\left( p \hspace{-.45em}/\hspace{.1em}_1-p
\hspace{-.45em}/\hspace{.1em}_4\right)\right]_{\sigma\mu}\left[
\gamma_5 \left( p \hspace{-.45em}/\hspace{.1em}_1-p
\hspace{-.45em}/\hspace{.1em}_4\right)\right]_{\lambda\nu}}{\left(
p_1-p_4\right)^2-M^2}\ \tau^{a}_{i_3 i_2}\tau^{a}_{i_4 i_1}\,.
\label{LOopeCov}
\end{eqnarray}
In order to simplify some technical aspects of solving the
leading-order equation we separate the pion denominator as
\begin{equation}\label{PiDen}
\frac{1}{k^2-M^2} = - \frac{1}{\vec k^2+M^2} +\left(
\frac{1}{k^2-M^2}+ \frac{1}{\vec k^2+M^2} \right)\,,
\end{equation}
and take the first term on the right-hand side of Eq.~(\ref{PiDen})
as the leading-order contribution in the effective
potential.\footnote{Without this simplification the Bethe-Salpeter
equation has to be solved, which is technically more involved.} The
second term is of higher order and is taken into account
perturbatively. The resulting OPE potential reads
\begin{eqnarray}
i \, V^{i_3 i_4,i_1 i_2}_{0,\pi; \lambda \sigma ,\mu \nu}  \left(
\vec p\,',\vec p \right) &=& \sum_{a=1}^3 \, i\,
\frac{g_A^2}{4\,F^2}\,\frac{\left[ \gamma_5 \left( p
\hspace{-.45em}/\hspace{.1em}_1- p
\hspace{-.45em}/\hspace{.1em}_3\right)\right]_{\lambda\mu}\left[
\gamma_5 \left( p \hspace{-.45em}/\hspace{.1em}_1-p
\hspace{-.45em}/\hspace{.1em}_3\right)\right]_{\sigma\nu}}{\left(
\vec p_1-\vec p_3\right)^2+M^2}\
\tau^{a}_{i_3 i_1}\tau^{a}_{i_4 i_2}\nonumber\\
&-&\sum_{a=1}^3\, i\, \frac{g_A^2}{4\,F^2}\,\frac{\left[ \gamma_5
\left( p \hspace{-.45em}/\hspace{.1em}_1-p
\hspace{-.45em}/\hspace{.1em}_4\right)\right]_{\sigma\mu}\left[
\gamma_5 \left( p \hspace{-.45em}/\hspace{.1em}_1-p
\hspace{-.45em}/\hspace{.1em}_4\right)\right]_{\lambda\nu}}{\left(
\vec p_1-\vec p_4\right)^2+M^2}\ \tau^{a}_{i_3 i_2}\tau^{a}_{i_4
i_1}\,. \label{LOopepot}
\end{eqnarray}
The projection reduces Eq.~(\ref{LOopepot}) to the following form,
\begin{eqnarray}
i\, \tilde V^{i_3 i_4,i_1 i_2}_{0 \pi; l s,m n}  \left( \vec p\,',
\vec p \right) &=& \sum_{a=1}^3 \,i\,
\frac{g_A^2}{4\,F^2}\,\frac{\left[ \vec\sigma\cdot \left(\vec p-\vec
p\,'\right)\right]_{lm}\left[ \vec\sigma\cdot \left(\vec p-\vec
p\,'\right)\right]_{sn}}{\left( \vec p-\vec p\,'\right)^2+M^2}\
\tau^{a}_{i_3 i_1}\tau^{a}_{i_4 i_2}\nonumber\\
&-&\sum_{a=1}^3 \, i\, \frac{g_A^2}{4\,F^2}\,\frac{\left[
\vec\sigma\cdot \left( \vec p+\vec p\,'\right)\right]_{sm}\left[\vec
\sigma\cdot \left( \vec p+\vec p\,'\right)\right]_{ln}}{\left( \vec
p+\vec p\,'\right)^2+M^2}\ \tau^{a}_{i_3 i_2}\tau^{a}_{i_4 i_1}\,.
\label{opepotreduced}
\end{eqnarray}

The leading-order scattering amplitude satisfies the equation
\begin{eqnarray}
\tilde T^{i_3 i_4,i_1 i_2}_{0; l s,m n}\left( p_3, p_4, p_1,p_2
\right)&=&\tilde V^{i_3 i_4,i_1 i_2}_{0; l s,m n}\left( \vec
p\,',\vec p\right) -\, i \int \frac{d^4 k}{(2\,\pi)^4} \ \tilde
V^{i_3 i_4,i j}_{0; l s,
a b}\left( \vec p\,',-\vec k\right) \nonumber\\
& \times & \,\frac{m^2\,F^2(L^2,\vec k\,{}^2)}{\left(\vec
k^2+m^2\right)\,\left( p_0-k_0 -\sqrt{\vec
k^2+m^2}+i\,\epsilon\right) \left( p_0+k_0-\sqrt{\vec
k^2+m^2}+i\,\epsilon\right)}\nonumber \\
&\times & \, \tilde T^{i j,i_1 i_2}_{0; a b,m n}\left( P/2
-k,P/2+k,p_1,p_2\right)\,, \label{MeqLo}
\end{eqnarray}
where
\begin{equation}\label{V0total}
   \tilde V_0 =\tilde  V_{0,C}+ \tilde V_{0,\pi}\,.
\end{equation}
As the leading-order potential does not depend on $k_0$, the
integration over this variable can be performed easily by
iterating Eq.~(\ref{MeqLo}), performing the $k_0$ integration in
each term and summing up the resulting terms. Doing so we obtain
\begin{eqnarray}
\tilde T^{i_3 i_4,i_1 i_2}_{0; l s, m n}\left( \vec p\,',\vec
p\right)&=&\tilde V^{i_3 i_4,i_1 i_2}_{0; l s, m n}\left( \vec
p\,',\vec p\right) - \frac{m^2}{2}\, \int \frac{d^3 k}{(2\,\pi)^3} \
\tilde V^{i_3 i_4,i j}_{0; l s, a b}\left( \vec p\,',-\vec k \right)
\nonumber\\
&\times& \frac{F^2(L^2,\vec k\,{}^2)}{\left(\vec k^2+m^2\right)\,
\left( p_0-\sqrt{\vec k^2+m^2}+i\,\epsilon\right)}\, \tilde T^{i
j,i_1 i_2}_{0; a b, m n}\left( -\vec k,\vec p\right)\,.
\label{MeqLOk0integrated}
\end{eqnarray}
\begin{figure}
\epsfig{file=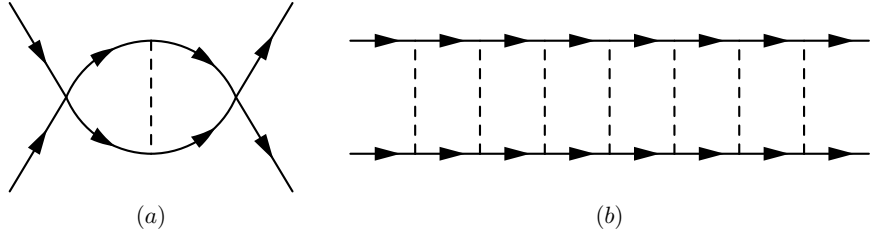,width=0.7\textwidth}
\caption[]{\label{nnampliter:fig} Two examples of iterations of
the LO equation.}
\end{figure}
Next we split the two-nucleon propagator into two parts,
\begin{eqnarray}
\frac{1}{\left( \vec k^2+m^2 \right)\left( p_0 - \sqrt{\vec
k^2+m^2}+i\,\epsilon\right)}
% = \frac{p_0 + \sqrt{\vec k^2+m^2}}{\left( \vec
%k^2+m^2 \right)\left( p_0^2 - \vec k^2-m^2\right)}\nonumber\\&& \\
& = & \frac{2\, m}{(\vec k^2+m^2)\left( \vec p\,^2 - \vec
k^2+i\,\epsilon\right)} \nonumber\\
& + & \frac{\left(p_0 + \sqrt{\vec k^2+m^2}\right)-2 \,m}{\left(
\vec k^2+m^2 \right) \left( \vec p\,^2 - \vec
k^2+i\,\epsilon\right)}, \label{2Npropagator}
\end{eqnarray}
and take the first term as the leading order propagator while
including the second part in perturbative corrections. The resulting
leading-order equation reads
\begin{eqnarray}
\tilde T^{i_3 i_4,i_1 i_2}_{0; l s, m n}\left( \vec p\,',\vec
p\right)&=&\tilde V^{i_3 i_4,i_1 i_2}_{0; l s,m n}\left( \vec
p\,',\vec p\right) - m^3\, \int \frac{d^3 k}{(2\,\pi)^3} \ \tilde
V^{i_3 i_4,i j}_{0; l s, a b}\left( \vec p\,',-\vec
k\right)\nonumber\\
&\times&  \,\frac{F^2(L^2,\vec k\,{}^2)}{\left(\vec k^2+m^2\right)\,
\left( \vec p\,^2 - \vec k^2+i\,\epsilon\right)}\, \tilde T^{i j,i_1
i_2}_{0; a b, m n}\left( -\vec k,\vec p\right)\,.
\label{MeqLOk0integratedLO}
\end{eqnarray}

Let us investigate the ''consistency problem'' which manifests
itself in the sensitivity of the scattering amplitude to the choice
of the parameter $L$. For that purpose we vary $L$ from $750$ MeV to
infinity. Divergences which occur in the limit $L\to\infty$ will be
referred to as UV divergences. Simple UV counting shows that
iterations of Eq.~(\ref{MeqLOk0integratedLO}) lead to finite
diagrams for $L\to\infty$. For example, the diagrams in
Fig.~\ref{nnampliter:fig} are finite. This differs from the standard
non-relativistic formulation, in which Fig.~\ref{nnampliter:fig} (a)
contains an overall logarithmic divergence $\sim M^2\, m^2$ and the
diagram in Fig.~\ref{nnampliter:fig} (b) contains a logarithmic
divergence $\sim (Q m)^6$. The divergences in these diagrams can be
absorbed in contact interaction terms of the second and sixth order
(and higher-order iterations require higher-order interaction terms,
up to infinity) \cite{Savage:1998vh}.

\begin{figure}
\epsfig{file=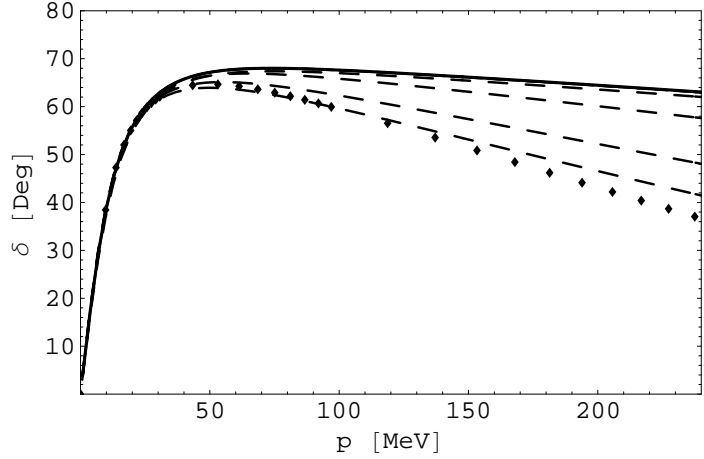,width=0.6\textwidth}
\caption[]{\label{nn1s0:fig} $^1S_0$ partial wave $np$ phase shifts
for different values of $L$ parameter compared with the data points
from the Nijmegen PWA \cite{Stoks:1993tb}. The curves are shown for
$L=$ 750, 1000, 2000, 6000, 20000, 30000, 100000 (MeV), the solid
line corresponding to the largest value.}
\end{figure}

\begin{figure}
\epsfig{file=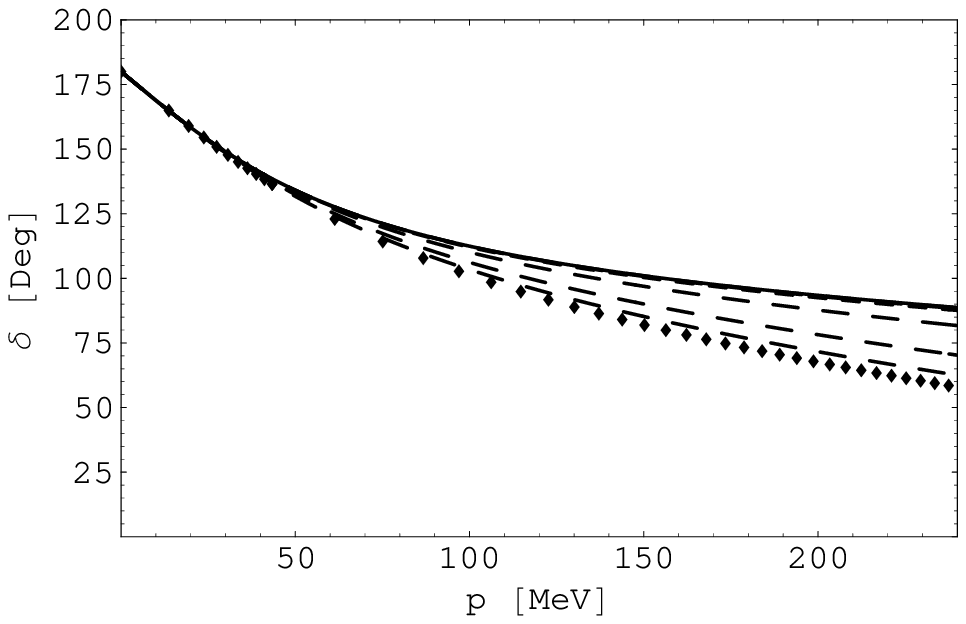,width=0.6\textwidth}
\caption[]{\label{nn3s1:fig} $^3S_1$ partial wave $np$ phase shifts
for different values of $L$ parameter compared with the data points
from the Nijmegen PWA \cite{Stoks:1993tb}. The curves are shown for
$L=$ 750, 1000, 2000, 6000,10000, 20000, 50000 (MeV), the solid line
corresponding to the largest value.}
\end{figure}

\begin{figure}
\epsfig{file=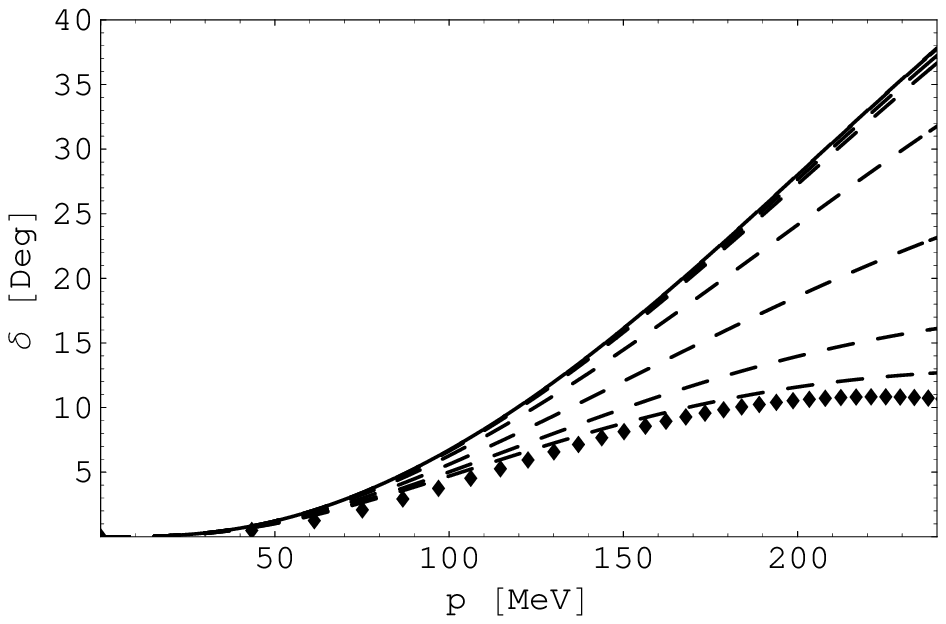,width=0.6\textwidth}
\caption[]{\label{nn3p0:fig} $^3P_0$ partial wave $np$ phase shifts
for different values of $L$ parameter compared with the data points
from the Nijmegen PWA \cite{Stoks:1993tb}. The curves are shown for
$L=$ 750, 1000, 2000, 6000, 20000, 30000, 100000, 500000 (MeV), the
solid line corresponding to the largest value.}
\end{figure}

Equation ~(\ref{MeqLOk0integratedLO}) can be rewritten in a more
familiar form. To that end we multiply
Eq.~(\ref{MeqLOk0integratedLO}) with
$$\frac{m^2\,F(L^2,\vec p\,'\,{}^2)\, F(L^2,\vec p\,{}^2)}{[(m^2+\vec
p\,^2)(m^2+\vec p\,'^2)]^{1/2}}
$$ and obtain
\begin{eqnarray}
{\cal T}^{i_3 i_4,i_1 i_2}_{0; l s, m n}\left( \vec p\,',\vec
p\right)&=& {\cal V}^{i_3 i_4,i_1 i_2}_{0;
l s,m n}\left( \vec p\,',\vec p\right) \nonumber\\
&-& m\, \int \frac{d^3 k}{(2\,\pi)^3} \ {\cal V}^{i_3 i_4,i j}_{0; l
s, a b}\left( \vec p\,',-\vec k\right) \, \frac{1}{ \vec p^2-\vec
k^2+i\,\epsilon} \,{\cal T}^{i j,i_1 i_2}_{0; a b,m n}\left( -\vec
k,\vec p\right)\,, \label{MeqLOk0integratedFinal}
\end{eqnarray}
where
\begin{eqnarray}
{\cal T}_{0}\left( \vec p\,',\vec p\right) &=& \frac{\tilde
T_{0}\left( \vec p\,',\vec p\right)\, F(L^2,\vec p\,'\,{}^2)\,
F(L^2,\vec p\,{}^2)
\,m^2}{[(m^2+\vec p\,'\,^2)(m^2+\vec p\,^2)]^{1/2}}\nonumber\\
{\cal V}_{0}\left( \vec p\,',\vec p\right) &=& \frac{\tilde
V_{0}\left( \vec p\,',\vec p\right)\,F(L^2,\vec p\,'\,{}^2)\,
F(L^2,\vec p\,{}^2)\,m^2}{[(m^2+\vec p\,'\,^2)(m^2+\vec
p\,^2)]^{1/2}}\,. \label{newcaldefinitions}
\end{eqnarray}
Note that the potential ${\cal V}_0(\vec p,\vec q)$ has a milder
ultraviolet behavior than the standard potential $\tilde V_{0}\left(
\vec p\,',\vec p\right)$ due to the denominator in
Eq.~(\ref{newcaldefinitions}). In the limit $L\to\infty$ the factors
$F(L^2,\vec p\,{}^2)$ and $F(L^2,\vec p\,'\,{}^2)$ approach $1$. The
OPE part of the effective potential ${\cal V}_0(\vec p\,',\vec p)$
of Eq.~(\ref{newcaldefinitions}) then behaves like $\sim 1/p^3$ for
large $p$ in the $^1S_0$ wave (after absorbing the local part in the
redefinition of the contact interaction term) and generates no
divergences (perturbatively as well as non-perturbatively). This is
in contrast to the OPE part of $\tilde V_{0}\left( \vec p\,',\vec
p\right)$ which goes as $\sim 1/p^2$. The most singular part of the
effective OPE potential in the $^3S_1$ wave requires a single
counter-term, which is present in the contact interaction part of
our LO potential. For higher partial waves the singular behavior of
the potential is screened by the angular momentum barrier and
therefore no counter-terms are required. Analyzing the convergence
criterium for Neumann series
%for the Fredholm integral equations of the second type
we have checked that the integral equations for higher partial waves
have unique solutions, i.e. the angular momentum barrier provides a
repulsion strong enough for our problem. The numerical analysis
confirms the above conclusions. Some of these phase shifts
calculated for different values of the parameter $L$ are shown in
Figs.~\ref{nn1s0:fig}, \ref{nn3s1:fig} and \ref{nn3p0:fig}. Note
that different choices of the parameter $L$ correspond to different
renormalization schemes. By choosing the renormalization scheme
properly one can improve the convergence of perturbative series for
physical quantities. Our figures suggest that the best description
of the phase shifts at leading order is obtained for $L\sim 750$
MeV. This indicates that the higher-order corrections are smaller
for this choice of the renormalization condition.

We conclude that all counter-terms required for the absorbtion of
the strong $L$ dependence in the LO equation
(\ref{MeqLOk0integrated}), perturbatively as well as
non-perturbatively,  are present in the LO potential. Therefore,
Eq.~(\ref{MeqLOk0integrated}) is free of the ''problems of
consistency of Weinberg's power counting'' encountered in solving
the corresponding Lippmann-Schwinger equation
\cite{Kaplan:1996xu,Kaplan:1998tg,Savage:1998vh}.\footnote{The
Lippmann-Schwinger equation corresponds to the leading-order term in
the $1/m$ expansion of Eq.~(\ref{MeqLOk0integrated}).}

Note that Eq.~(\ref{MeqLOk0integrated}) in comparison with the LS
equation contains the re-summation of an infinite number of
higher-order terms. In the authors' opinion both approaches are
equivalent and self-consistent, provided that one includes
compensating terms in the effective Lagrangian and chooses the
cutoff parameter appropriately in the LS approach. Therefore we
expect the numerical results for physical quantities in our
approach to be comparable to that of the cutoff LS equation
approach \cite{Epelbaum:2005pn}.

\subsection{\label{contactinteraction}Higher-order Corrections}

Corrections to the leading-order amplitude are calculated
perturbatively and to any finite order can be expressed in a closed
form as a sum of a finite number of terms analogous to
Eq.~(\ref{NLOeqsolutionS}). These expressions can be interpreted as
sums of a finite number of Feynmann diagrams where the leading-order
amplitude $T_0$ is also interpreted as an NN vertex. The
straightforward analysis of
Eqs.~(\ref{MeqLOk0integratedLO})-(\ref{MeqLOk0integratedFinal})
shows that (for $L\to\infty$) $T_0(\vec p,\vec p')$ cannot diverge
stronger than $\sim {\rm ln} p$ ($\sim {\rm ln} p'$) for
$p\to\infty$ ($p'\to\infty$). This UV behavior of $T_0$ guarantees
that all divergences generated in the limit $L\to\infty$ in Feynman
diagrams contributing to corrections at any finite order can be
canceled by counter-term contributions also present at the given
order. As a result the new semi-relativistic approach suggested in
this work is free of ''consistency problems'' inherent in the
standard non-relativistic approach.

\section{\label{conclusions}Summary}

We consider a new approach to the nucleon-nucleon scattering problem
in baryon chiral perturbation theory. Starting with a
Lorentz-invariant form of the effective Lagrangian with higher
derivative terms we work out a new framework where the
nucleon-nucleon scattering amplitude is calculated order-by-order in
small parameters like the pion mass and nucleon three-momenta. We
develop a systematic power counting for the effective potential
which is defined as the sum of all two-nucleon irreducible diagrams.
The leading-order amplitude is obtained by solving an equation with
the LO potential consisting of the contact interaction plus OPE.
Higher-order corrections to the potential as well as to the
two-nucleon propagator are taken into account perturbatively. The
parameter $L$, introduced through terms with higher derivatives,
parameterizes the renormalization scheme dependence of the
scattering amplitude present in our calculations. Due to the UV
behavior of our semi-relativistic two-nucleon propagator,
analogously to the KSW approach, the strong $L$-dependence of
physical quantities to any finite order is absorbed in the
redefinition of a finite number of parameters of the effective
potential at given order. On the other hand, analogously to
Weinberg's power counting, the OPE potential is of leading order and
is treated non-perturbatively. Our new approach preserves all
underlying symmetries to the given order of accuracy and the
self-consistency of the power counting is explicit in this new
formulation.

\acknowledgments

D.D., J.G. and M.R.S.~acknowledge the support of the Deutsche
Forschungsgemeinschaft (SFB 443).

\end{document}